\documentclass[journal]{vgtc}                

\ifpdf
  \pdfoutput=1\relax                   
  \usepackage{graphicx}                
  \DeclareGraphicsExtensions{.pdf,.png,.jpg,.jpeg} 
\else
  \usepackage{graphicx}                
  \DeclareGraphicsExtensions{.eps}     
\fi%

\graphicspath{{figures/}{pictures/}{images/}{./}} 

\usepackage{microtype}                 
\PassOptionsToPackage{warn}{textcomp}  
\usepackage{textcomp}                   
\usepackage{mathptmx}                  
\usepackage{times}                     
\usepackage{cite}                      
\usepackage{tabu}                      
\usepackage{booktabs}
\usepackage[dvipsnames]{xcolor}
\usepackage{paralist}
\usepackage{caption}
\usepackage{xspace}
\usepackage{algorithm}
\usepackage{algpseudocode}
\usepackage{amsmath}
\usepackage{multirow}
\usepackage{float}

\definecolor{myblue}{RGB}{52, 152, 219}
\definecolor{myorange}{RGB}{230, 126, 34}
\definecolor{mygreen}{RGB}{46, 204, 113}
\definecolor{cluster1}{RGB}{179, 205, 227}
\definecolor{cluster2}{RGB}{251, 180, 174}
\definecolor{cluster3}{RGB}{204, 235, 197}
\definecolor{cluster4}{RGB}{222, 203, 228}
\definecolor{cluster5}{RGB}{254, 217, 166}
\usepackage{enumitem}

\newcommand{\name}{GraphFederator\xspace}
\newcommand{\model}{FGRM\xspace}


\title{\name: Federated Visual Analysis for Multi-party Graphs}

\author{Dongming Han, Wei Chen, Rusheng Pan, Yijing Liu, Jiehui Zhou, Ying Xu, Tianye Zhang,\\ Changjie Fan, Jianrong Tao, and Xiaolong (Luke) Zhang}
\authorfooter{
\item
Dongming Han, Wei Chen, Rusheng Pan, Yijing Liu, Jiehui Zhou, Ying Xu, and Tianye Zhang are with the State Key Lab of CAD \& CG, Zhejiang University, Hangzhou, Zhejiang 310058, China. E-mail: dongminghan@zju.edu.cn
\item
Changjie Fan and Jianrong Tao are with NetEase Fuxi AI Lab, Hangzhou, China. E-mail: fanchangjie@corp.netease.com, hztaojianrong@corp.netease.com
\item
Xiaolong (Luke) Zhang is with Penn State, University Park, Pennsylvania, United States. E-mail: lzhang@ist.psu.edu
\item  Corresponding authors: Wei Chen. E-mail: chenvis@zju.edu.cn
}

\abstract{
This paper presents \name, a novel approach to construct joint representations of multi-party graphs and supports privacy-preserving visual analysis of graphs. Inspired by the concept of federated learning, we reformulate the analysis of multi-party graphs into a
decentralization process. The new federation framework consists of a shared module that is responsible for joint modeling and analysis, and
a set of local modules that run on respective graph data. Specifically, we propose a federated graph representation model (FGRM) that is learned
from encrypted characteristics of multi-party graphs in local modules. We also design multiple visualization views for joint visualization, exploration,
and analysis of multi-party graphs. Experimental results with two datasets demonstrate the effectiveness of our approach.} 

\teaser{
  \centering
  \includegraphics[width=\linewidth]{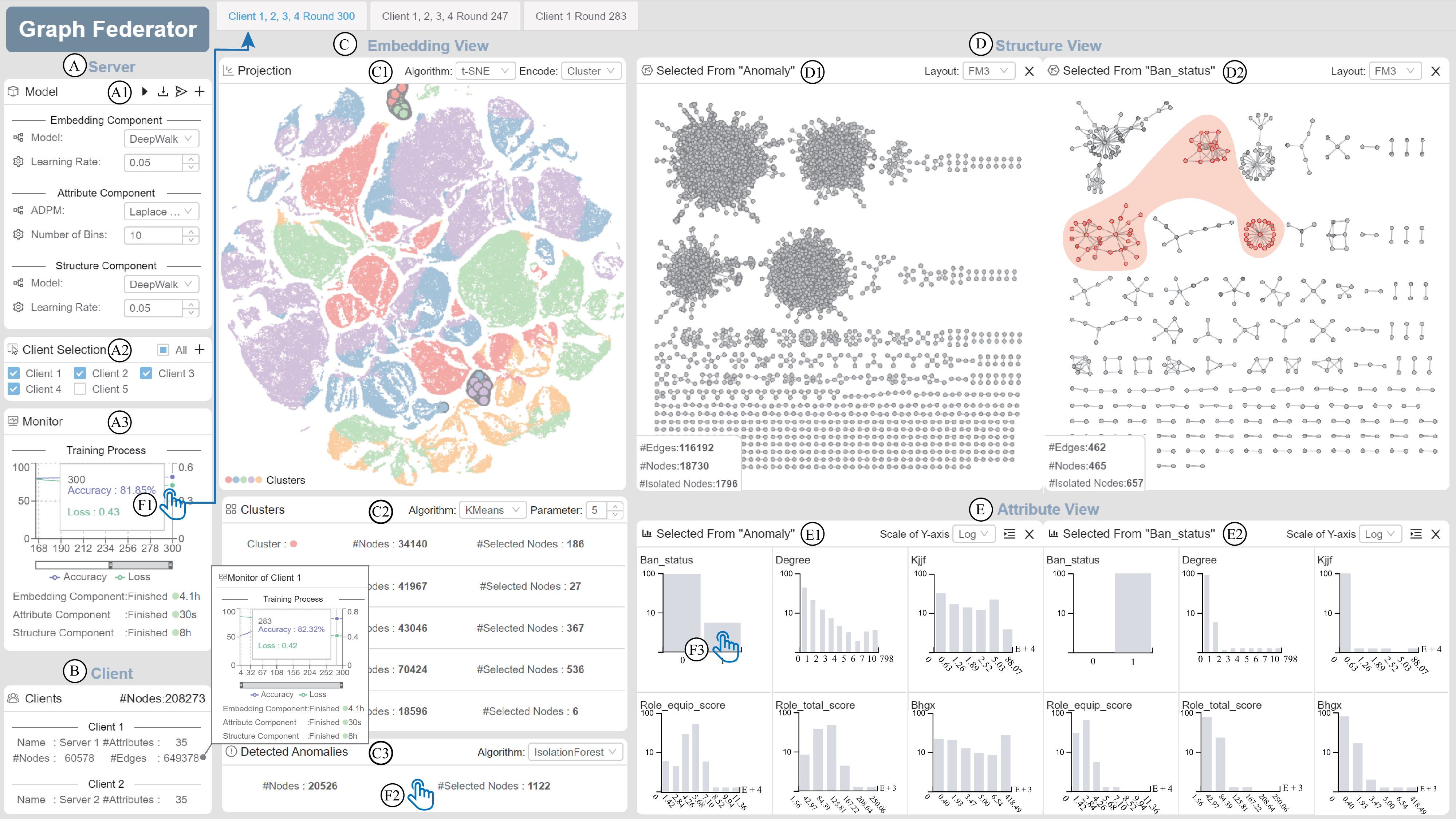}
  \caption{Illustration of GraphFederator with case study 1 (\autoref{section:case1}). The server view (A) supports model configuration (A1), client selection (A2), and process monitoring (A3). The client view (B) shows the information of selected clients. The user selects federated graph representations of multi-party graphs (F1) or a single client from the training process. The representations are visualized in other views (C) (D1) (E1). The embedding view (C), consists of the projection view (C1), the cluster view (C2), and the anomaly view (C3). Federated structure representations are shown in the structure view (D) by using a node-link diagram (D1). The attribute view (E) employs histograms (E1) to visualize  federated attribute representations. When detected anomalous players are selected (F2), their federated graph representations are visualized in other views (D1) (E1). The filtering condition of banned status is set from the corresponding bin of the histogram (F3), and federated graph representations of banned players from anomalous players are shown in other views (D2) (E2). The structure view (D2) shows links among players. It indicates that some banned players know each other and may belong to a studio that controls many plug-in players’ accounts.
  }
	\label{fig:teaser}
}
\vgtcinsertpkg

\begin{document}

\maketitle

\section{Introduction} \label{intro}
Visual analysis of multi-party graphs plays an important role in helping us understand real-world complex data ~\cite{von2011visual,wang2018visual,cao2015g}, such as ego-network analysis in social media~\cite{wu2016survey, zhao2016egocentric}, disease diagnosis in healthcare~\cite{liu2016graph}, and anomaly detection in public security~\cite{cao2015targetvue,zhang2017survey}. Various features or models extracted from multi-party graphs can be integrated to support a comprehensive understanding of the entire graph data.
Using the integrated information, we can conduct more comprehensive investigations.
For instance, by combining knowledge graphs of patients and diseases from multiple hospitals, doctors can gain a deeper understanding of diseases and develop best treatment plans.

One main bottleneck of exploiting multi-party graphs is data accessibility. Early studies on graph visual analysis assume that the graph data is freely accessible. Currently, however, more and more graph data are distributed (e.g., on servers in different organizations). To analyze such data, we need to combine multi-party graphs and examine them as an entirety. Considering of privacy and security, raw data in distributed clients may be prohibited from accessing. This leads to two challenges for visual analysis of multi-party graphs. The first is the federation of multi-party graphs.
Because raw data should be kept locally, creating a joint representation of data in all clients must resort to privacy-preserving feature extraction techniques. This situation is even exaggerated by the fact that features of different graphs may be different. A uniformed joint representation that is capable of characterizing the essential features of each graph is needed.
The second challenge is that the federated analysis based on joint representations is difficult. Designing a generalized framework for federating various and complex analysis tasks remains a huge challenge.

Visual analysis of multi-party graphs based on the joint representations requires new approaches. Conventional graph analysis, which is performed in a centralized model, or only uses limited accessible graph data, cannot be applied to multi-party graph analysis.
Existing solutions for distributed graph analysis~\cite{hong2015pgx,mccune2015thinking} mainly focus on the partitions of data and analysis tasks for the purposes of performance improvement, and are incapable of supporting multi-party graph analysis either.

One way to support privacy-preserving decentralized graph analysis is to build a decentralized federation of both data features and the analysis. We propose \name, a novel federation approach that constructs joint representations of multi-party graphs, and supports privacy-preserving visual analysis of graphs.
Inspired by federated learning, we reformulate the analysis of multi-party graphs into a decentralization process. The new federation framework consists of a shared module that is responsible for joint modeling and analysis, and a set of local modules that run on respective graph data.
Specifically, we propose a federated representation model
that is iteratively learned from the encrypted characteristics of multi-party graphs in local modules.
We design multiple visualization views for joint visualization, exploration, and analysis of multi-party graphs.
The contributions of this paper include:
\begin{itemize}[parsep=0pt, topsep=3pt]
    \item a federated graph representation model to represent and extract distinctive features from multi-party graphs; and
\item a federated visual analysis approach to support privacy-preserving graphs analysis.
\end{itemize}

The rest of this paper is organized as follows. Related work is discussed in Section 2.
Section 3 introduces the problem formulation.
Section 4 explains our design goals and the overview of our approach.
Our federated graph representation model is introduced in Section 5.
Section 6 presents the visual interface.
The evaluation is discussed in Section 7.
Discussions and the conclusion are given in Section 9 and Section 10, respectively.

\section{Related Work} \label{related}
\subsection{Visual Analysis of Graph Data}
A graph analysis task is usually defined as the analysis of entities or associated properties~\cite{pretorius2014tasks}. Here, entities denote nodes, links, paths, and networks, while the properties include structures and derived features.
Graph analysis tasks~\cite{lee2006task} can be classified into four groups: topology-based tasks, attribute-based tasks, browsing tasks, and overview tasks. Complex tasks can be decomposed into a set of basic tasks. Alternatively, tasks can be represented as a combination of two fundamental tasks~\cite{nobre2019state}: analyzing topology for given attributes, and analyzing attributes for a given topological structure. These two tasks are supported with respect to topological structures, including nodes, edges, clusters, node neighbors, paths, and substructures.

A recent study~\cite{brehmer2013multi} proposes a multi-level typology to facilitate specific task classifications. Likewise, 29 group-level graph visualization tasks~\cite{saket2014group} are classified into four groups: group only tasks, group-node tasks, group-link tasks, and group-network tasks. From the viewpoint of graph-based sensemaking, four categories of graph visualization tasks~\cite{pienta2015scalable} are introduced: visualization and exploration; global, local, and hybrid views; subgraph mining and interaction. Without loss of generality, this paper follows their notations and design specified tasks. A typical graph analysis system named Network Repository~\cite{rossi2015network} provides users with the ability to explore, visualize, and compare data along many different dimensions interactively and in real-time. By combining global network statistics, local node-level network statistics, and features, users can easily discover key insights into the data.

\subsection{Distributed Analysis and Federated Learning}

Machine learning is benefited from the ability to train increasingly sophisticated models with the unprecedented growth of data collection~\cite{2019arXiv191209789V}. To overcome the problem of high computational cost for analyzing large-scale data, parallel or distributed computing has become popular~\cite{canini2012sibyl}. Similarly, decentralized machine learning approaches can be reformulated from centralized versions. There are two basic building blocks of distributed learning algorithms: matrix multiplication and data shuffling~\cite{lee2017speeding}.
Much effort is paid to improve communication efficiency. For instance, minimizing the number of rounds of communication works well for cases where data is unevenly distributed over an extremely large number of nodes~\cite{konevcny2016federated}. A new framework is proposed to manage asynchronous data communications between clients and servers with flexible consistency, elastic scalability, and fault tolerance~\cite{li2013parameter}.

To protect the privacy data during the process of communication, a Homomorphic Encryption (HE) scheme~\cite{rivest1978data} is designed to preserve structures of the original message space. HE can be leveraged to privacy-preserving training or prediction of linear regression, linear classifiers, decision trees, matrix factorization, and neural networks. To address the problem of non-linear activation functions, a HE-based neural network scheme~\cite{orlandi2007oblivious} is proposed with an interactive protocol between the data owner and the model owner. The calculated transformation of the data owner is swapped in an encrypted form with the result from the model.

As a pioneering work on privacy-preserving computing, secure multi-party computation (MPC)~\cite{goldreich1998secure} guarantees that clients can only get the final cumulative model weight. Recently, federated learning emerged as a new privacy-protection scheme to construct a global machine learning model with distributed multiple clients~\cite{mugunthan2020privacyfl,yang2019federated,konevcny2016federated,konevcny2016federated1,mcmahan2016federated, bagdasaryan2018backdoor}.
To improve the training efficiency while protecting the privacy of multiple parties, local training data is kept from the central server.
It is trained in a decentralized manner on multiple remote clients without transferring raw data. By integrating parameters from clients, the server can compose a global model.
For large-scale graph data, distributed processing is needed. A distributed implementation of the Dijkstra algorithm~\cite{chandy1982distributed} can handle various graph problems like the depth first search in an undirected graph. $TUX^2$~\cite{xiao2017tux2} is a distributed engine that supports the layout and computation of graphs with billions of edges and outperforms other state-of-the-art graph engines by using a series of graph-based optimizations.

To our best knowledge, research on using federated learning for graph analysis is rare. A distributed learning algorithm on graph generalizes the previous work on federated learning~\cite{lalitha2019peer} and provides a fully decentralized framework with localized data of individual nodes kept from one another. The entire learning process over nodes does not use a central server and hence is a peer-to-peer method. A distributed graph neural network is constructed by following the scheme of federated learning~\cite{mei2019sgnn}. The algorithm uses a similarity matrix to capture the high-distance structure of nodes precisely in graph neural network.



\section{Problem Formulation}
For the reason of data privacy, directly analyzing graphs in multiple clients is difficult.
Our solution is to extract joint representations from multi-party graphs and use joint representations for analysis.
However, existing federated learning frameworks are inapplicable for multi-party graphs, because the data characteristics and analysis tasks of multi-party graphs are quite diverse.
Designing a model that extracts joint representations with privacy-persevering is a nontrivial issue.

Given $K$ parties, we define $G_k$ as a graph in the $k$-th party, where $G_k=(V_k,E_k)$ and $V_k, E_k$ represent the node set and edge set, respectively.
Each graph $G_k$ owns the identical attributes $A_k={a_1, a_2,..,a_J}$, where $J$ denotes the dimension of attributes.
Each node $v$ has a public ID.
Nodes from different graphs may have the same public ID.
Our goal is to construct a federated graph representation model that generates the joint representations $R$ of nodes $V_k\in G_k$.
Sensitive information should be kept locally: attributes of each node, such as age, year and salary; links among nodes; and personal privacy information, like name, e-mail, and address.

\section{Approach Overview}

\subsection{Design Goals} \label{sec-design-goal}
We work closely with five domain experts. Two of them are professors whose research focus is on federated learning and privacy analysis, respectively. The other three include one professor and two Ph.D. students. Their research interests are all related to graph representation learning.
We also consulted two experts of an online game publisher and provider. Both of them have experience in federated learning and graph analysis research.
Through discussions with these experts, we identify the following design goals:

\begin{compactitem}
\item[\textbf{\textcolor{myblue}{G}}] \textbf{\textcolor{myblue}{G}lobal representation learning model for multi-party graphs data.}\label{g:G}
Considering that the data distribution can be distinctive for different parties, a globalized criterion is needed to ensure all computations are secured and generating satisfying results.

\item[\textbf{\textcolor{mygreen}{R}}] \textbf{\textcolor{mygreen}{R}epresentations learning for each party graph.}
The main goal of multi-party graphs representation learning is to derive the localized graph representation of each party. In particular, localized representations should be:
\begin{compactenum}[\begingroup
    \color{mygreen} {R}1
    \endgroup]
\item \textit{Privacy-preserving:}\label{g:L1}
Parties are not allowed to share or transmit their raw data, which should be kept locally to avoid the leakage of sensitive information.

\item \textit{High-quality:}\label{g:L2}
Local representations learned by our decentralized scheme should achieve comparable performance with the ones learned in a centralized manner.

\item \textit{Information-diverse:}\label{g:L3}
Learned local representations are comprehensive and can enlighten insights about graphs from various aspects. Such representations should be extracted on the basis of rich and diverse characteristics of a graph, such as the structure information, the side information (node attributes), and graph embedding results.
\end{compactenum}

\item[\textbf{\textcolor{myorange}{C}}]\textbf{\textcolor{myorange}{C}ustomizations support.}\label{g:C}
Analysis tasks vary with different users, and thus a flexible scheme is needed. Users should be allowed to customize analysis methods and control relevant parameters of different steps upon their preferences.
\end{compactitem}

\subsection{Approach Overview}
\begin{figure*}[htbp]
    \includegraphics[width=\linewidth]{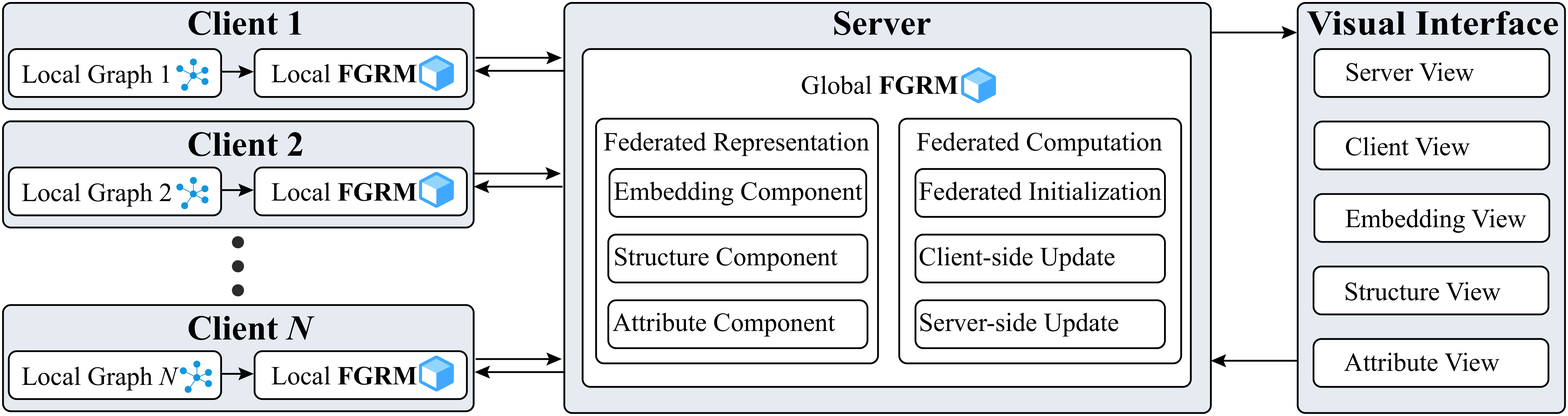}
    \caption{Overview of our approach.
        The server distributes \model to clients.
        Each client updates \model and sends graph representations to the server.
        The server collects and handles graph representations and sends them back to each client.
        After a certain number of iterations, federated graph representations of multi-party graphs are generated.
        The visual interface is used to configure \model and illustrate federated graph representations.
    }
    \label{fig:overview}
\end{figure*}

Our general approach is shown in \autoref{fig:overview}. The central component of our approach is a server that runs a global federated graph representation model (FGRM). The server communicates with individual clients, each of which owns its local graph data and runs its local \model. And with the user interface to provide data for various views.

\model is designed to extract multiple graph representations (\textbf{\textcolor{myblue}{G}}). It runs in a server-client mode,
and contains two components: graph representation and federated computing.

The graph representation of each local graph is computed based on three components (\textbf{\textcolor{mygreen}{R2}}, \textbf{\textcolor{mygreen}{R3}}): the embedding component, the structure component, and the attribute component.
The federated computing is used to federate graph representations from multi-party graphs.

The federated computing (\textbf{\textcolor{mygreen}{R1}}, \textbf{\textcolor{myorange}{C}}) includes three parts: federated initiation, client-side update, and server-side update.
In the federated initiation, the server distributes \model, predefined encryption schemes, and related rules to multiple remote clients.
The client-side module updates \model with the local graph and sends extracted graph representations to the server.
The server-side module collects and handles these representations.
Next, the server sends back the representations to each client.
The update process is iterated until the specified number of rounds is reached.
Finally, the server receives federated graph representations from multi-party graphs.

We design and implement a visual interface to visualize different federated graph representations.
Users configure and build \model by custom schemes in the server view for extracting federated representations from multi-party graphs (\textbf{\textcolor{myorange}{C}}).
The embedding view, the attribute view, and the structure view display the federated embedding representation, the federated attribute representation, and the federated structure representation, respectively.

\section{Federated Graph Representation Model}

\subsection{Graph Representation}
\model supports the construction of three different types of graph representations: graph embedding, node attribute, and structure information. These representations depict graph information from different aspects, making \model suitable for various visual analysis tasks.
For the purpose of simplicity, the federated graph representations are denoted as $R=(W_{emb}, W_{att}, W_{struc})$. Here, the federated embedding representation is denoted as $W_{emb}$. The federated atrribute representation is denoted as $W_{att} = (W_{att_1}, W_{att_2},..., W_{att_P})$, and $P$ denotes the dimension of attributes. The federated structure representation is denoted as $W_{struc}$.

\subsubsection{The Embedding Component}
The embedding component is used to construct a graph embedding representation of a graph. This representation converts the node set into low dimensional vectors in a canonical space~\cite{cai2018comprehensive}.

\textbf{Input:}
The input includes a graph $G=(V,E)$ with attributes $A$, an embedding model, and corresponding parameters.

\textbf{Extraction:}
For a graph $G$, its embeddings are generated from a graph embedding model.
Some models only use the topology to extracted representations, while some require node features.
For embedding learning models that require topology and features as the input, the final representation is extracted directly by the model with $G$ and $A$.
For those models that only require topology as the input, basic embedding vectors are extracted by the model with $G$.
Then, features of nodes are extracted by following these steps.
First, one-hot vectors are extracted to represent categorical attributes and text attributes.
Then, normalized vectors are extracted to represent the numerical attributes.
Feature vectors are then concatenated by these two vectors.
Thereafter, feature vectors are reduced to the same number of dimensions as basic embedding vectors.
The embedding representation is concatenated by reduced feature vectors and basic embedding vectors, yielding a high dimensional vector for each node.

\textbf{Output:}
A set of vectors of nodes is generated.

\subsubsection{The Structure Component}
The structure component is used to construct the connection relationships of a graph.
The raw connection relationships are sensitive.
Therefore, the structures are restructured from the embeddings of the graph.

\textbf{Input:}
The input includes a graph $G=(V,E)$ with $A$, an embedding method, and corresponding parameters.

\textbf{Extraction:}
The embedding can be extracted by the embedding component or by specified embedding models.
Then, the distance matrix is calculated by node embeddings.
The edges among nodes are extracted by the user-specified reconstruction method with the distance matrix.
The reconstruction is controlled by predefined parameters to avoid privacy disclosure.
Here, different embedding methods can be used to support different analysis tasks.
Some methods are strong in link prediction and graph reconstruction, while some are good at node clustering.
This component and embedding component can use different embedding methods for specified analysis tasks.

\textbf{Output:}
Reconstructed graph structures are output.

\subsubsection{The Attribute Component}
The attribute component is used to construct the attribute distribution of a graph. Attribute distributions are extracted to avoid privacy disclosure.

\textbf{Input:}
The input includes a graph $G=(V, E)$ with attributes $A$, the bin size of the attribute distribution and filter condition.

\textbf{Extraction:}
For numeric attributes, the bin size of the distribution is specified by users.
For categorical attributes, the bin size of the distribution is the dimensions of attributes.
Those data types that are less meaningful for counting (e.g., \textit{name}) are not used.
Moreover, the topology attributes can also be extracted. These attributes are important to the analysis of the nodes of a graph, such as identifying the social influence in the social media network.
To avoid privacy exposure, the extraction only constructs topology distributions rather than topology values.
The distribution of the topology attribute of one node characterizes the node.
The supported topology metrics include:
(1) \textit{Degree}, (2) \textit{Betweenness}, (3) \textit{Eigenvector}, (4) \textit{PageRank}, (5) \textit{Clustering Coefficient}, (6) \textit{Average Nearest Neighbors Degree (KNN)}.
This component will filter out specific nodes according to the filter conditions, and count the attribute distribution of these nodes (e.g., extracting the attribute distribution of players whose age is between 10 and 30).

\textbf{Output:}
The attribute distribution of each node is extracted.

Note that other graph representation components can also be designed for specified graph visual analysis tasks.

\subsection{Federated Computation}
Federated computation is used to federate graphs and generate joint representations.
The process contains three steps: federated initialization, server-side update, and client-side update.

\subsubsection{Federated Initialization}
Federated initialization sets the configuration of the federation model: encryption schemes, rules for computing feature vectors, and the distribution of the model.

\textbf{Encryption:}
We provide different encryption schemes to protect the transmission of data, although the transmitted data contains no sensitive information.
Sometimes, the data owner still has concerns about data privacy, so our model employs encrypted federated average and encrypted model training~\cite{yang2019federated}.
Specifically, our model is implemented with the federated functions of TensorFlow (e.g., \textit{federated\_mean} and \textit{federated\_max}).
TensorFlow uses homomorphic encryption~\cite{rivest1978data} to secure the transmission process.

Attributes of nodes can be used to specify an individual's identity uniquely. A traditional way of protecting privacy is to transfer only attribute distributions of nodes. Unfortunately, the identity can be re-identified by exploiting the side information or schematic meaning of data~\cite{sweeney2000simple,wang2017utility}. Thus, \model employs multiple attribute distribution protection models to improve results:
(1) syntactic anonymization
models: $k$-anonymity~\cite{sweeney2002k} and $l$-diversity~\cite{machanavajjhala2007diversity}~\cite{li2007t}.
(2) differential privacy models: Laplace mechanism~\cite{dwork2006calibrating} and exponential mechanism~\cite{mcsherry2007mechanism}.

\textbf{Computing feature vectors:}
The server counts fields of categorical attributes from all clients,
and sets the corresponding one-hot vector for each field.
Then, the server counts the maximum and minimum values of each numerical attribute and formulate the normalization standard of each attribute.
Finally, each client calculates feature vectors of each node based on computing rules.

\textbf{The distribution of \model:}
The server distributes our \model and relevant settings to each client.
The weights of the model (the hidden layer) are the embedding results of the graph.
The weights compose an $N \times M$ matrix, where the row number $N$ is the number of different nodes of all graphs, and the column number $M$ is the dimensions of the vector of a node.
$M$ can be configured by users.
A row presents the embedding representation of a node.
Models in both clients and the server have the same weight.
However, the nodes of local graphs are different.
To find the embedding of the row corresponding to each node, the server unifies the row index of each node of all clients.
It should be noted that the node counts of graphs in clients are different.
The model makes statistics of the number of graph nodes from all clients and the server, and then sets the index of the row of the matrix corresponding to each node.

\subsubsection{Client-side Update}
Client-side update runs on each client.
For the embedding component and the structure component, the server distributes a graph representation model and the initial weights of the model to each client in the federated initiation.
Each client executes the model with an initial weight.
Then, each client learns the graph representation from the learning model with the local graph per round,
and calculates the gradients that encode the differences between weight pairs between two weights.
Thereafter, each client sends gradients to the server.
This process does not transmit the raw data, but only transfers the gradients of the learning model.

Each client calculates $W_{att}$ by a user-specified attribute distribution protection model.
Then, $W_{att}$ is encrypted and transmitted to the server by means of TensorFlow.
Note that, secure aggregation protocols and homomorphic encryption algorithms~\cite{bonawitz2017practical} are supported by TensorFlow.

\begin{algorithm}[htbp]
    \caption{The update in the $k$th client}
    \label{alg:client}
    \begin{algorithmic}[1]
        \Require
        $W$: weights sent by the server-side,
        $\eta$: the learning rate of the learning model;
        $Model$: the embedding model.
        \Ensure
        $\Delta$: gradients of \model,
        $n^k$ : the updated weight of the client.
        \Function {ClientUpdating}{$W, Model, \eta$}
        \State $W \gets \Call{ServerExecutes}{ }$
        \State $\beta \gets $ (Local data is divided into minibatches)
        \State $n^k \gets |\beta|$ // \textit{The updated weight is used to do the weighted average}
        \State $W_{init} \gets W$
        \For{batch $b \in \beta$}
        \State $W \gets W-\eta \nabla Model(W;b)$
        \EndFor
        \State $\Delta^k \gets W - W_{init}$

        \State \Return{$\Delta^k , n^k$}
        \EndFunction

    \end{algorithmic}
\end{algorithm}

\begin{algorithm}[htbp]
    \caption{The attribute extraction in the $k$th client}
    \label{alg:client}
    \begin{algorithmic}[1]
        \Require
        $ADPM$: the attribute distribution protection model.
        \Ensure
        $E_{att_k}$: the encrypted attribute representation.
        \Function {clientAttribute}{$ADPM$}
        \State $W_{att_k} \gets ADPM(G_k,A_k)$
        \State $E_{att_k} \gets \Call{ttf.federated.map}{W_{att_k}}$ // \textit{ TensorFlow transmits it to the server based on encryption schemes.}
        \State \Return{$E_{att_k}$}
        \EndFunction
    \end{algorithmic}
\end{algorithm}

\subsubsection{Server-side Update}
The server randomly generates the weights of the embedding learning model and distributes the model with weights to each client.
The server collects gradients of models from clients to fulfill federated average per round.
Then the server computes the weighted average of gradients according to the node number of each client.
Next, new weights are computed based on weights and averaged gradients, and are sent back to each client.
The server executes the weights updating process iteratively until the specified number of rounds is reached.
The weights of the model, $W_{emb}$, are learned from graphs in clients.
This transmission process does not transmit raw data.

The federated structure representations $W_{struc}$ is generated based on $W_{emb}$ with link prediction and graph restructuration algorithms. For instance, the distance matrix of nodes is calculated.
Edges can be generated by calculating the nearest $|E|$ node pairs.
The federated attribute representation $W_{att}$ is calculated by means of TensorFlow.

\begin{algorithm}[htbp]
    \caption{Server-side update clients per round.}
    \label{alg:server}
    \begin{algorithmic}[1]
        \Require
        $K$: the amount of clients sampled per update,
        $T$: the iteration count,
        $ADPM$: the attribute distribution protection model,
        $M_{emb}$: the embedding model of the embedding component,
        $\eta_{M_{emb}}$: the learning rate of  $M_{emb}$,
        $M_{struc}$: the embedding model of the structure component,
        $\eta_{M_{struc}}$: the learning rate of $M_{struc}$.
        \Ensure
        $W_{emb}$: the federated embedding representation of \model,
        $W_{att}$: the federated attribute representation of \model,
        $W_{struc}$: the federated structure representation of \model.
        \Function {ServerExecutes}{$K, T, M_{emb}, M_{struc}$}
        \State $W_{emb} \gets \Call{GetEmbedding}{K, T, M_{emb}, \eta_{M_{emb}}}$
        \State $W_{M_{struc}} \gets \Call{GetEmbedding}{K, T, M_{struc}, \eta_{M_{struc}}}$
        \State $W_{struc} \gets \Call{Restructure}{W_{M_{struc}}}$
        \For{each client $k=1,2,..., K$}
        \State $E_{att_k} \gets \Call{clientAttribute}{ADPM}$
        \EndFor
        \State $W_{att} \gets \Call{ttf.federated.mean}{{E_{att_1}, E_{att_2},...,E_{att_K}}}$ // \textit{TensorFlow calculates the average distributions of clients based on encryption schemes.}
        \State \Return$W_{emb}, W_{att}, W_{struc}$
        \EndFunction

        \Function {GetEmbedding}{$K, T, Model, \eta$}
        \State initialize $W_1$ randomly.
        \For{each round $t=1,2,..., T$}
        \State $(\Delta_t^k, n_t^k) \gets \Call{ClientUpdating}{W_t, Model, \eta}$ from the $k$th client
        \State $\overline{W}_{t} \gets \sum_k\Delta^k_t $ // \textit{Sum the weighted average}
        \State $\overline{n}_t \gets \sum_k{N_t^k}$
        \State$\Delta_t \gets \frac{\overline{W}_{t}}{\overline{n}_t} $ //\textit{Updating the weighted average based on $F$}
        \State $W_{t+1} \gets W_{t} + \Delta_t$
        \EndFor
        \State \Return$W$
        \EndFunction
    \end{algorithmic}
\end{algorithm}
\section{Visual Interface}
\autoref{fig:teaser} shows the interface of our system. It consists of five views:
a server view (\autoref{fig:teaser} (A)) that provides \model configuration (\autoref{fig:teaser} (A1)), data selection (\autoref{fig:teaser} (A2)) and model monitoring (\autoref{fig:teaser} (A3));
a client view (\autoref{fig:teaser} (B)) that shows the information and the process state of clients;
an embedding view (\autoref{fig:teaser} (C)) that shows federated embedding representations;
a structure view (\autoref{fig:teaser} (D)) that shows federated structure representations: and an attribute view (\autoref{fig:teaser} (E)) that visualizes federated attribute representations.

\subsection{The Server View}
Within the server view (\autoref{fig:teaser} (A)),
users can configure graph learning models, multiple parameters and encryption schemes (\autoref{fig:teaser} (A1)), and monitor the running process of \model (\autoref{fig:teaser} (A3)).
The loss and the accuracy of \model are shown in a line chart (\autoref{fig:teaser} (A3)).
Users can choose the federated graph representation anytime in a training process and visualize them in other views (\autoref{fig:teaser} (F1)).
Through these visual graphs, users can explore the training representation, verify the accuracy of the representation, and evaluate the models and parameters.

\subsection{The Client View}
In the client view (\autoref{fig:teaser} (B)), users can observe the general information, the running state, and the training process of each client.
Users can select a client to examine its process state and visualize its federated graph representations in other views.

\subsection{The Embedding View}
The embedding view contains (\autoref{fig:teaser} (C)) multiple visualization components associated with federated graph embedding representations.
Following visual graphs are supported: the projection view (\autoref{fig:teaser} (C1)), the clustering view (\autoref{fig:teaser} (C2)), and the anomaly view ((\autoref{fig:teaser} (C3))).

\textbf{Projection view:}
This view shows the distance of nodes of multi-party graphs in high-dimensional embedding space.
Each point in the view represents a node (\autoref{fig:teaser} (C1)).
The node color encodes clustering information or anomaly information.
Interactive actions such as panning, zooming, and lasso-based selection are supported.
Users can select nodes to study attribute distributions in the attribute view and their reconstructed structures in the structure view.

\textbf{Cluster view:}
This view uses a table to visualize the clustering result of nodes from multi-client graphs according to federated embedding representations(\autoref{fig:teaser} (C2)).
Users can select nodes from one of the clusters for visualizing their attribute distributions and reconstructed structures in other views (\autoref{fig:teaser} (D2) (E2)).

\textbf{Anomaly view}:
This view lists the anomaly detection results of multi-party graphs based on federated embedding representations (\autoref{fig:teaser} (C3)).
When users select some nodes, their attribute distributions and reconstructed structures are shown in other views.

\textbf{Control panel:}
With the control panel, users can configure parameters or methods in each view (in the top right corner).
In the projection view, different projection methods including MDS~\cite{cox2008multidimensional}, and $t$-SNE~\cite{maaten2008visualizing} can be chosen.
In the cluster view, different node clustering algorithms and corresponding parameters, including $K$-Means~\cite{krishna1999genetic} and DBSCAN~\cite{ester1996density} can be set.
In the anomaly view, One-Class SVM~\cite{chen2001one} or IsolationForest~\cite{liu2008isolation} can be used for nodes' anomaly detection.

\subsection{The Structure View}
The structure view (\autoref{fig:teaser} (D)) depicts reconstructed structures of selected nodes from other views.
To make structure exploration clear and avoiding heavy overlapping, this view hides nodes without any edge.
Different layout methods in the view can be chosen.

\subsection{The Attribute View}
In the attribute view (\autoref{fig:teaser} (E)), multiple histograms are used to visualize the attribute distribution of multi-party graphs.
Users can interact with bins of histograms (\autoref{fig:teaser} (F3)) to specify attribute distributions in the corresponding attribute interval.
The \textit{y}-axis in each histogram can be either linear or logarithmic.

\section{Evaluation}
In implementing \name, the front-end client was developed with the React framework and D3.js. Our in-house graph visualization engine is employed with rich user interactions, and flexible customizations.
TensorFlow is used to compute the federated average of the attribute component. Pytorch is used to execute the graph embedding learning model.

\textbf{Datasets.}
\textit{DBLP Dataset:}\label{dataset-dblp}
This is a paper citation graph dataset.
Each node represents a paper, and an edge represents the citation relationship between two papers, which has 5 attributes.
Papers and their citation graphs in four areas are extracted to form a graph: AI, System, Theory, and Interdisciplinary. The total number of papers in each area are 64,232, 62,020, 14,430, and 62,708, respectively.

\textit{NetEase-Game-Player Dataset (NEGP):}\label{dataset-netease}
This is a game player transaction graph dataset provided by NetEase Co.\footnote{http://game.163.com} It is collected in five servers of a massively multi-player online role playing game.
In each graph, a node represents a player, and an edge represents a transaction between two players.
Each player has 36 attributes related to the player's role information and account status, such as \textit{role\_level}, \textit{role\_class} and \textit{create\_date}.
\begin{table}[htbp]
    \centering
    \begin{tabular}{|c|c|c|c|c|}
        \hline
        Dataset                                        & Client ID & \#Nodes & \#Edges   & \#Attributes        \\ \hline
        \multirow{5}{*}{DBLP}                          & Year 2014 & 33,171  & 63,212    & \multirow{5}{*}{5}  \\ \cline{2-4}
                                                       & Year 2015 & 38,374  & 74,632    &                     \\ \cline{2-4}
                                                       & Year 2016 & 40,755  & 89,831    &                     \\ \cline{2-4}
                                                       & Year 2017 & 38,541  & 84,338    &                     \\ \cline{2-4}
                                                       & Year 2018 & 32,356  & 84,257    &                     \\ \hline
        \multirow{5}{*}{NEGP} & Game server 1  & 60,578  & 649,378   & \multirow{5}{*}{36} \\ \cline{2-4}
                                                       & Game server 2  & 45,687  & 378,350   &                     \\ \cline{2-4}
                                                       & Game server 3  & 40,657  & 363,080   &                     \\ \cline{2-4}
                                                       & Game server 4  & 61,351  & 579,937   &                     \\ \cline{2-4}
                                                       & Game server 5  & 72,043  & 587,594   &                     \\ \hline
    \end{tabular}
    \caption{The data profiles.}\label{tab:data}
\end{table}

\subsection{Experiments}
We conducted several experiments to evaluate our approach.
Experiments were conducted in a PC with a single Intel (R) Xeon (R) Gold 5218 CPU (a basic frequency of 2.3GHz and 16 cores), 128GB internal memory, and single RTX 2080TI GPU.

\subsubsection{Federated Embedding Representation}\label{sec-fml-evaluate}
Considering that the federated scheme may have a lower quality of embedding representations, the performance and computing cost with our approach and its centralized counterpart were tested on two datasets DBLP~\cite{tang2008arnetminer} and NEGP.

\textbf{Configurations.}
We ran three configurations to evaluate the performance of different models: embedding learning in a single client (ELSC), centralized embedding learning (CEL), and our federated embedding component (FEC).

\textbf{Data processing.}
\textit{DBLP Dataset~\cite{tang2008arnetminer}:}
The citation graphs of years 2014-2018 were used in our experiments with CEL and FEC, each of which located in a client.
The abstract of each paper was encoded as a fix-length vector based on word frequency count.
For ELSC, the embedding generated from the graph of the year 2016, which is the largest graph among them, was used to generate the performance baseline (ELSC).
The sub-filed information was employed as the ground truth to evaluate embedding representations with three configurations.

\textit{NEGP Dataset}:
The game player transaction graphs from five servers were evaluated with CEL and FEC.
The graph from game server 1 was used to derive ELSC.
Players were classified into two categories: high and low levels of in-game consumption.
The consumption information was used as the ground truth to evaluate embedding representations with three configurations.

\textbf{Settings.}
Two popular embedding learning methods were employed: DeepWalk~\cite{perozzi2014deepwalk}, an unsupervised approach for learning node representation without utilizing attribute vectors; and GAT~\cite{velivckovic2017graph}, one of the most popular neural graph architectures of graphs, which captures both the structure and the feature of each node.
The random walk step of DeepWalk was firstly applied to generate samples by setting the number of walk times to be 80, and the length of walks to be 40. DeepWalk learned node representations by using the skip-gram model. We set the dimension of representation and the length of the sliding window to be 128 and 10, respectively. The GAT model was configured as follows: the number of layers was 3; the intermediate dimension of representations was 256; three heads were used in GAT. To train GAT, we employed the stochastic gradient descent (SGD) algorithm. The learning rate and the coefficient of L2 regularization were set to be 0.001 and 0.0001, respectively. DeepWalk and GAT of all clients were updated with 300 rounds.

We tested three configurations on the DBLP and NEGP datasets.
GAT has high computational complexity and memory consumption.
The CEL was not conducted on GAT.
All evaluations were derived by a 5-fold cross-validation.

\begin{figure*}[h]
    \includegraphics[width=\linewidth]{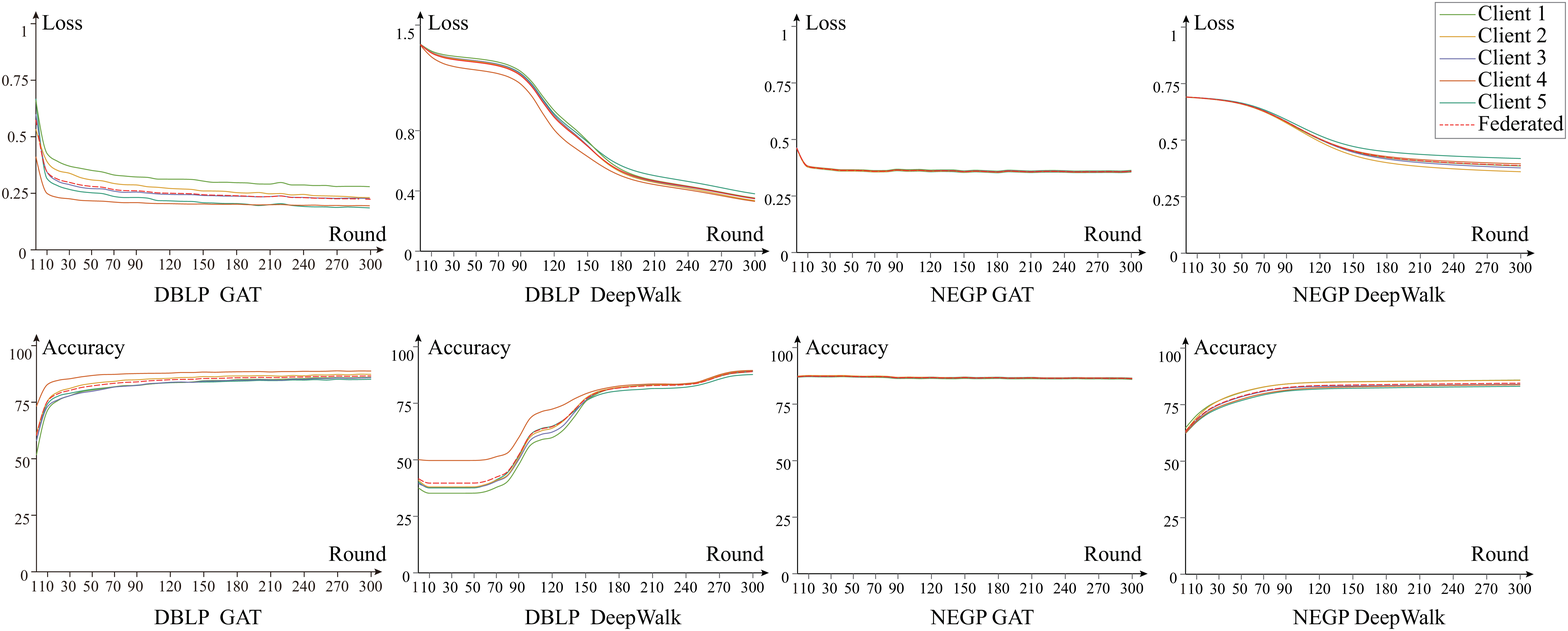}
    \caption{
        The loss and the accuracy of each client of federated embedding representations with FEC in each client on the DBLP and NEGP datasets, with two embedding learning models: GAT and DeepWalk.
    }
    \label{fig:eva-lines}
\end{figure*}

\begin{figure}[h]
    \includegraphics[width=\linewidth]{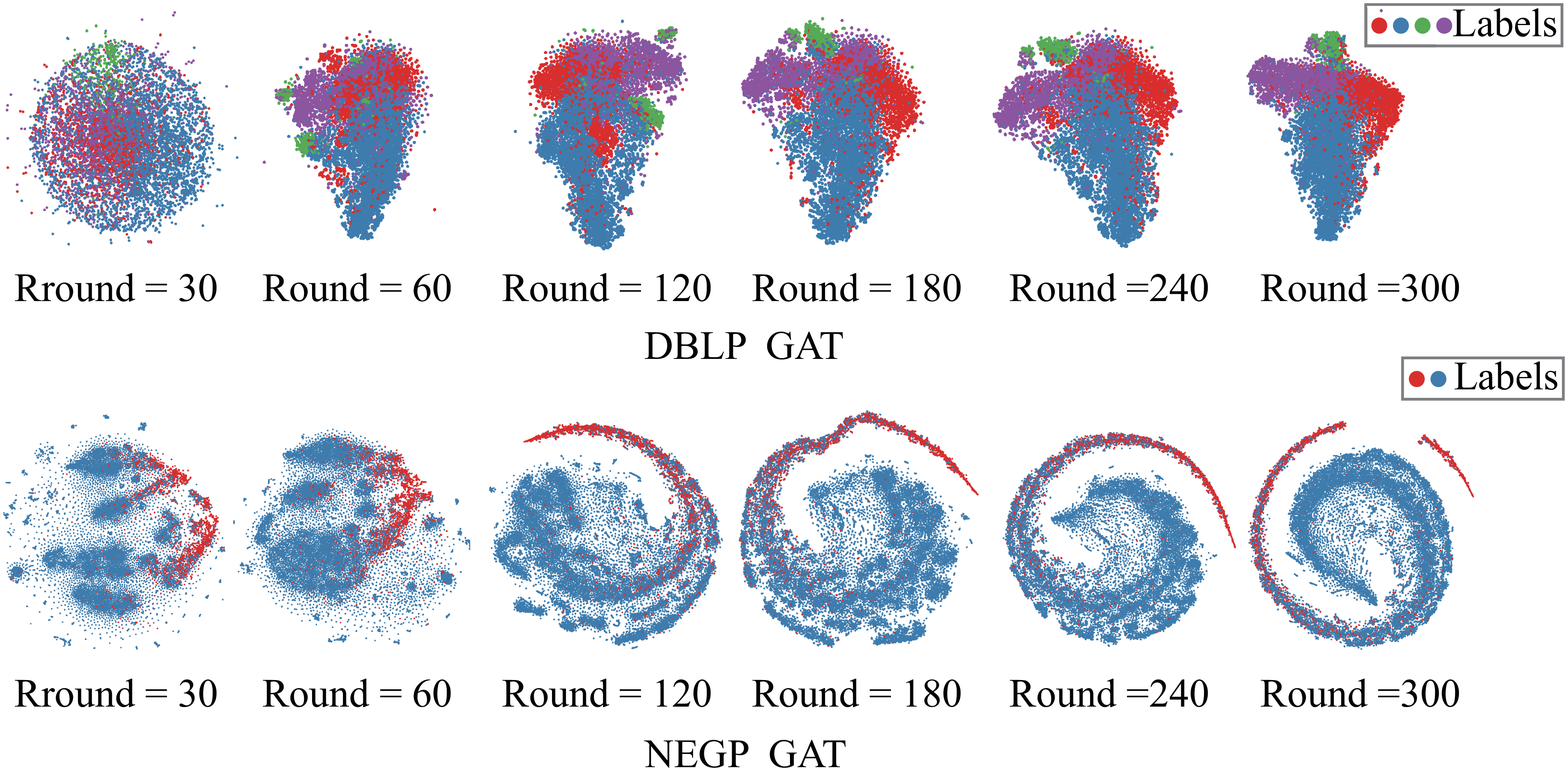}
    \caption{
        Projections (\textit{t}-SNE) of federated embedding representations with FEC in different training rounds on test datasets of DBLP and NEGP with GAT.
        The color encodes ground truth labels.
    }
    \label{fig:eva-projection}
\end{figure}

\textbf{Results.}
The accuracy and time consumption are shown in~\autoref{tab:learning}.
The accuracy of FEC has a similar performance with that of CEL, and the accuracy of FEC or CEL is better than that of SCEL. The time consumption of FEC is more than that of SCEL, and less than that of CEL.
Line charts of the loss of our approach are shown in ~\autoref{fig:eva-lines}.
Projections of test dataset embeddings extracted by our approach are shown in~\autoref{fig:eva-projection} by using \textit{t}-SNE.
In the first half of the training process, the embedding already has good adequate performance. Users can early terminate the training by observing the visualization of the results in real-time.

\begin{table}[]
    \centering
    \begin{tabular}{|c|c|c|c|c|}
        \hline
        Dataset                                    & Model                   & Configuration & Accuracy       & Time          \\
        \hline
        \multirow{6}*{DBLP}                        & \multirow{3}*{DeepWalk} & ELSC          & 89.04          & 2.3h          \\
                                                   &                         & CEL           & 91.87          & 8.1h          \\
                                                   &                         & FEC& 90.13 & {4.4h} \\
        \cline{2-5}
                                                   & \multirow{3}*{GAT}      & ELSC          & 86.74          & 0.4h          \\
                                                   &                         & CEL           & -              & -             \\
                                                   &                         & FEC  & 88.35 & 1.1h \\
        \hline
        \multirow{6}{*}{NEGP} & \multirow{3}*{DeepWalk} & ELSC          & 85.26          & 2.6h          \\
                                                   &                         & CEL           & 85.62          & 8.7h          \\
                                                   &                         & FEC  & 85.86 & 5.1h \\
        \cline{2-5}
                                                   & \multirow{3}*{GAT}      & ELSC          & 86.11          & 0.9h          \\
                                                   &                         & CEL           & -              & -             \\
                                                   &                         & FEC  & 87.59 & 1.7h\\
        \hline
    \end{tabular}
    \caption{The performance (in hours) with three configurations on the DBLP and NEGP datasets, with two embedding learning models: DeepWalk and GAT.}\label{tab:learning}
\end{table}

\subsubsection{Federated Attribute Representation}\label{federated-averag-sec}
We tested the computing cost of the federated average in the attribute component.
For the attribute protection model, the time consumption is very short.
The federated average supported by TensorFlow was used to encrypt attributes distributions of all clients.
The time consuming of the encrypt could not be ignored.

\textbf{Settings.}
The computing cost of three aspects was collected: the number of clients (1-30), the number of attributes (1-30), and the node number in each client (1-50,000).
We evaluated one aspect while fixing the other two.
Each evaluation was repeated for five times.

\begin{figure}[h]
    \includegraphics[width=\linewidth]{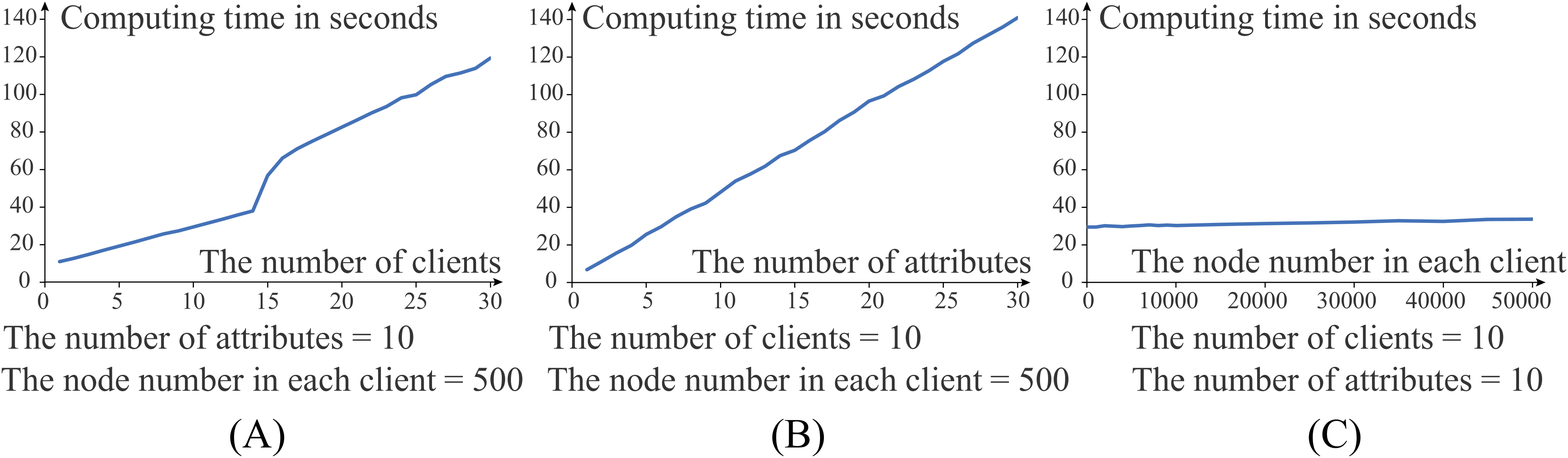}
    \caption{The computing cost of creating federated attribute representations.
        The computing cost increases linearly with the number of clients (A) and the number of attributes (B), but has little relevance with the node number in each client (C).
    }
    \label{fig:performance-average}
\end{figure}

\textbf{Results.}
The influence of each variable on the computing cost is reported in~\autoref{fig:performance-average}. The computing cost of the extaraction of federated attribute representations shows a linear complexity over the number of clients and attributes (\autoref{fig:performance-average} (A) (B)) but has little relevance with the node number in each client (\autoref{fig:performance-average} (C)).

\subsubsection{Federated Structure Representation}\label{test: structure}
We evaluated the performance of federated structure representation by link prediction evaluation metrics: the Area Under Curve (AUC) score and Precision~\cite{lu2011link}. Deepwalk was selected because it precisely captures the linkages among nodes from the node sequence generated by random walk. Supervised models such as GAT seek to minimize difference among intra-class nodes, and may lead to a result that node embeddings of same category are concentrative in space, which is problematic for linkage reconstruction.

\textbf{Setting.}
We used training representations of the entire dataset to test the performance.
We selected 10,000,000 edge pairs to test AUC score and set $L$ of Precision as 1000.

\textbf{Results.}
Given $G=(V,E)$ and $W_{emb}$, the distance matrix of nodes is calculated, and the time complexity is $O(|V|^2)$.
Edges can be generated by calculating the nearest $|E|$ node pairs. Its time complexity is $O(|V|^2log(|E|))$ by using the heap sorting technique.
The total time complexity $O(|V|^2 +|V|^2log(|E|))$.
Results are show in~\autoref{tab:structure}.
AUC score is high on two datasets.
Precision is not satisfactory on the NEGP dataset, probably because the dataset has a high data complicacy.

\begin{table}[htbp]
\centering
    \begin{tabular}{|c|c|c|c|}
        \hline
        Dataset & Model    & AUC      & Precision     \\ \hline
        DBLP    & DeepWalk &  99.89   & 94.7           \\ \hline
        NEGP    & DeepWalk &  98.98   & 76.5           \\ \hline
    \end{tabular}
\caption{The performance of federated structure representations for the DBLP and NEGP datasets.}\label{tab:structure}
\end{table}

\begin{figure*}[htbp]
    \includegraphics[width=\linewidth]{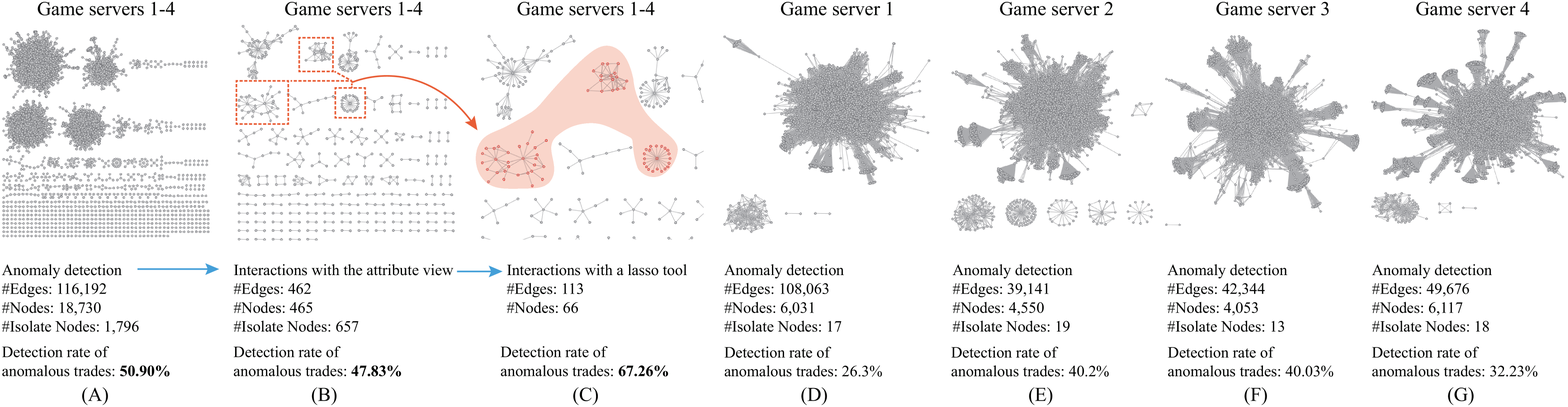}
    \caption{
        Anomaly detection with \name for the NEGP dataset. (A): reconstructed structures of anomaly detection by using federated graph representations from game servers 1-4; (B): reconstructed structures are interactively modified in the attribute view; (C): the selected structures are highlighted with a lasso tool; (D)-(G): raw structures of anomaly detection by using the graph representation from a single game server. By checking with the ground truth, detection rates of anomalous trades of \model detection and \name with interactions are much higher than only using a single game server graph.
    }
    \label{fig:case1-anomaly}
\end{figure*}

\subsection{Case Study}

\subsubsection{Case 1: NetEase-Game-Player Dataset}\label{section:case1}
We invited an expert to use our \name to analyze the game data. He works in NetEase and is skilled at game data analysis. We introduced our system and showed how it works. Then, he used \name to analyze and explore NEGP datasets freely. His interest was in verifying the validity of the learned model and anomalous trades with \name.
He took graphs from four different game servers as the input data (clients) of \model in the server view (\autoref{fig:teaser} (A2)).
Then, he configured \model to build federated graph representations (\autoref{fig:teaser} (A1)).
The general information of the graph from each client is shown in the client view (\autoref{fig:teaser} (B)).
The running status and the progress of \model are shown in the monitoring view (\autoref{fig:teaser} (A3)).
The monitor information indicates that the model runs well.

\begin{figure}[htbp]
    \includegraphics[width=\linewidth]{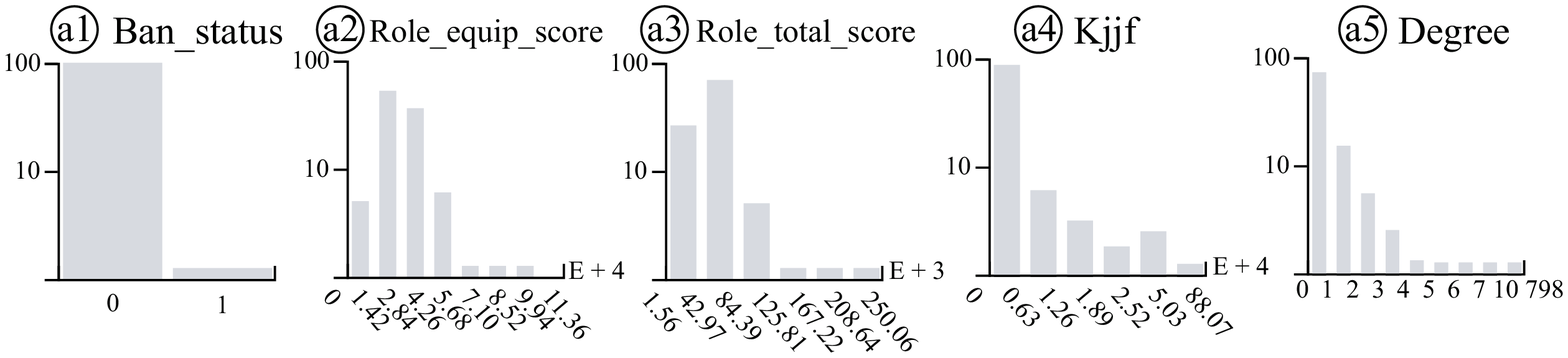}
    \caption{
        Attribute distributions of the NEGP dataset.
    }
    \label{fig:case1}
\end{figure}

With the obtained representations of all components after 300 rounds of training, the expert saw clear clusters of the embedding in the projection view (\autoref{fig:teaser} (C1)).
The structure view showed clear and diverse structures.
He concluded that \model performed well.
He studied the distributions of the main attributes of players (\autoref{fig:case1} (a1-5)) and found that the distributions of several attributes are interesting.
While most \textit{Role\_total\_score} (players' scores) are low, and a small number of players have scores distributed in the highest interval.
This result indicates that a small number of players' game progress leads to the vast majority.
The distribution of \textit{Role\_equip\_score} (players' equipment score) and  the distribution of \textit{Degree} (the number of trades of a player) are generally in a relatively low range (\autoref{fig:case1} (a2) (a5)), and imply that most players have low equipment scores and trading records.
These distributions helped the expert better understand the upgrading speed of the game, the consumption preferences of the players, and the preferences of players in different activities.

He selected 20,526 detected anomalous players (\autoref{fig:teaser} (F2)) to study anomalies in in-game trading behavior from the anomaly view.
He found that many anomalous players had no trading record in the structure view (\autoref{fig:teaser} (D1)), and many players' accounts were banned (\autoref{fig:teaser} (E1)) in the attribute view. Apparently, these players behaved differently from the rest of all players (\autoref{fig:case1} (a1)).
It should be noted that the status of being banned is one of the most important attributes for game analysis because it indicates that a player may have an illegal plugin or other behaviors that violate game fairness policies~\cite{tao2019mvan,tao2018nguard}.
He selected banned players from the corresponding histogram (\autoref{fig:teaser} (F3)), and the structure view showed trades among 1,122 players who were banned and detected as anomalies (\autoref{fig:teaser} (D2)).
There are 462 trades among 465 players.
It indicates that some banned players knew each other and may belong to a studio that controls the accounts of many plugin players. In addition, these players may accumulate their virtual wealth in the game onto a specific player's account.
He selected three structures (\autoref{fig:teaser} (D2)) because he believed that the selected structures were typical anomalous trading patterns based on his domain knowledge.
The short and intensive trading chain is the unique characteristic of anomalous trades.

The \textit{Degree} distribution showed that some banned players made many trades (\autoref{fig:teaser} (E2)).
Interestingly, he found that \textit{kjjf} (cumulative score of an activity) distributions were distributed in the lowest interval (\autoref{fig:teaser} (E2)), compared with those of all players (\autoref{fig:case1} (a4)).
He gave two possible explanations. The first one is that banned players were banned earlier, and they had no chance to accumulate scores. The second is that some plugin players seek to make money or upgrade the level.
They made no contribution to \textit{kjjf}, which is an indicator of entertainment activities.

The three structures with 66 players and 168 trades (\autoref{fig:case1-anomaly} (C)) were also presented to the expert. 113 trades are anomalous trades. The expert was surprised that our system achieved a high detection rate of anomalous trades.
Raw graph data of each game server was used to learn embeddings and detect anomalous trades. We found that trades among these players are very intensive (\autoref{fig:case1-anomaly} (D)-(G)).
We recorded structures of each step analysis in this case study.
These results were presented to the expert.
Using \model to detect anomalous trades from four game servers, yields detection rate of 50.90\% (\autoref{fig:case1-anomaly} (A)).
However, when a single game server data is employed, the detection rate are much lower (\autoref{fig:case1-anomaly} (D)-(G)).
This indicates that \textbf{federated graph representations extracted from multi-party graphs can capture more underlying insights}.

In this case study, the expert achieved a detection rate of 67.26\% (\autoref{fig:case1-anomaly} (C)).
\name empowers him to improve the detection rate of anomalous trades and discover anomaly trading patterns.

\begin{figure*}[htbp]
    \includegraphics[width=\linewidth]{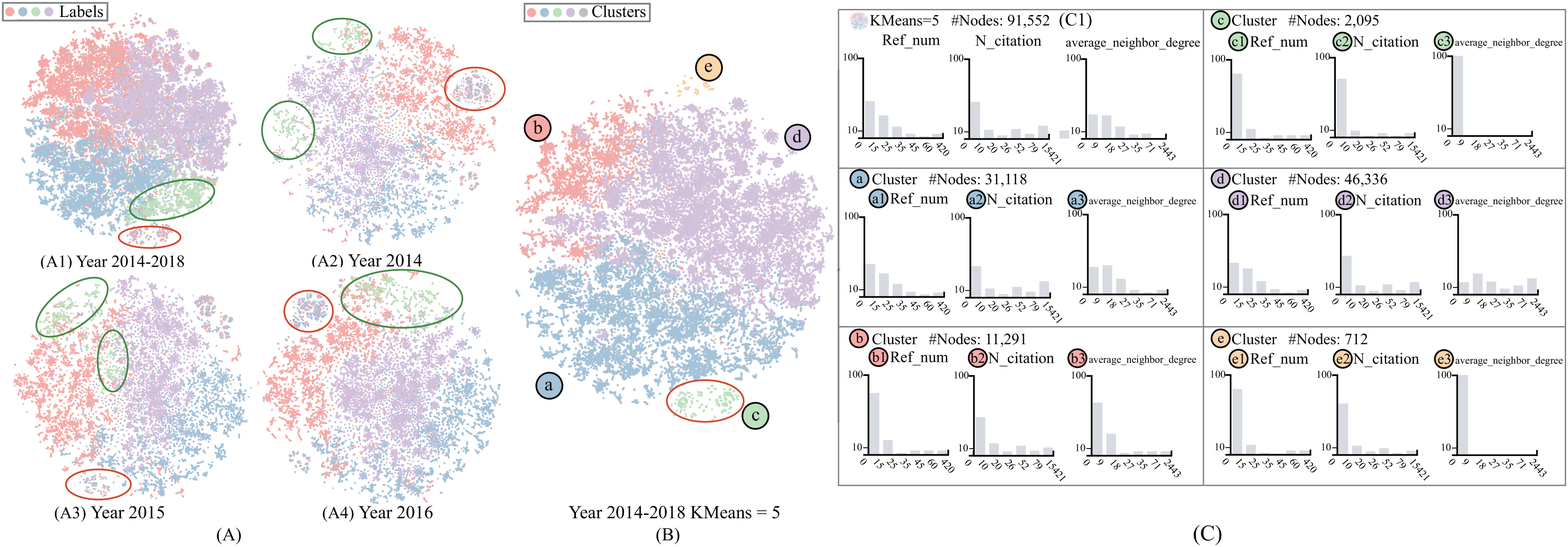}
    \caption{
        Clustering results with \name for the DBLP dataset case study.
        (A1) and (B) show the same projection of federated embedding representations on the DBLP dataset in 2014-2018 with DeepWalk.
        (A2), (A3) and (A4) show projections of representations by using the data of 2014, 2015, 2016, respectively.
        Colors of projections in (A) encode 4 ground truth labels. Colors in (B) encode 5 clusters calculated by the $K$-Means algorithm.
        (C) shows attribute distributions of the entire set (C1) and clusters.
        Areas of green and red circles show distributions of selected papers from different representations extracted by different data.
        }
    \label{fig:case2}
\vspace{-3mm}
\end{figure*}
\subsubsection{Case 2: DBLP Dataset}\label{section:case2}
We invited a professor to use our \name to analyze the DBLP dataset.
His research area is soical network analysis.
He wanted to analyze the clustering result of the paper citation data.

He chose data from 2014 to 2018 and configured the embedding component with DeepWalk.
He selected the training results of the last round and observed federated graph representations in other views.
He modulated different parameters of the $K$-Means algorithm from 3 to 6 and selected 5 (\autoref{fig:case2} (B)).
Then, he selected different clusters.
By observing the attribute view (\autoref{fig:case2} (C)), he found differences of attributes: \textit{Ref\_num} (the number of references), \textit{N\_citation} (the number of citations), and \textit{Average\_neighbor\_degree}, the distributions of which are shown in~\autoref{fig:case2} (C1).
Clusters (a) and (d) have similar distributions with distributions of the entire set of papers.
He inferred that most papers belong to these two clusters, so they have similar distributions.
However, the distributions of three clusters (b) (c) and (d) lie in the lower interval.
This indicates that those papers falling into these clusters are rarely cited.
He inferred that these papers were published recently.
The distributions in clusters (c) and (e) lie in the lowest interval.
Clusters (c) and (e) have fewer papers than other clusters.
The \textit{Ref\_num} and \textit{N\_citation} distributions indicate that the numbers of citations and references of papers in these two clusters are small.
The \textit{Average\_neighbor\_degree} implies that the number of citations and references of cited papers and reference papers is also small.
He concluded that papers in two clusters are not attractive.

We used raw data of each year from 2014-2016 to learn embeddings(\autoref{fig:case2} (A2)-(A4)).
\autoref{fig:case2} (A1) shows the projection of federated embedding representations of years 2014-2016.
The color encodes ground truth labels.
We found that papers with the green label are separated in 2014 and 2015, but clustered together in 2016.
Papers in red circles (\autoref{fig:case2} (A1)-(A4)) are clustered in terms of both data representations.
It is regarded as a cluster, even though they have different labels.
It indicates that federated graph representations extracted from multi-party graphs can capture more features and information, and help find clusters with unique features.

\subsection{Expert Reviews}\label{section:expert-review}


We interviewed the expert involved in our first case study (\autoref{section:case1}). The expert thought that the findings could help to improve strategies of the anomalous player detection.
He also believed that the clustering result of players of different clients might be used to stimulate more strategies for studying players of different clusters with unique features.
The professor in the second case study (\autoref{section:case2}) suggested that our system supports the extraction and visualizations of keywords or abstracts of papers. He hoped that analysts should be able to control privacy standards and analyze more data or dimensions; otherwise, some interesting insights would be missed.

To evaluate the effectiveness of our approach, we also conducted one-on-one interviews with five additional domain experts from NetEase Co.
They are all skilled at game data analysis, graph analysis, and federated learning.
With a live, hands-on demonstration for approximately 10 minutes, we showed them case studies. We discussed the feasibility of \model and findings and solicited feedback from them.
They all confirmed that our approach could help them analyze features and information of players without touching raw data, and \name empowers them to explore multiple aspects and features of graphs.
They liked our intuitive user interface for visual analysis of large-scale graphs. One expert commented on our system by saying: ``\emph{...The system can help me validate training results of the federated model by the visual interface and help accomplish various analysis tasks of multiple graphs.}
''
Another expert claimed: ``\emph{...When I want to analyze graphs which are distributed in multiple clients, the accessibility of graphs limits me. \model can solve these problems and give wonderful visualizations of features from graphs.}''


\section{Discussions}

\textbf{Privacy.}
In the computing of federated representations, the server computes the weighted average of gradients of each client model and sends weights to each client.
By jointly averaging gradients, the model can be trained by using multi-party graphs without switching raw data.
The federated average algorithm computes the attribute representations from each client graph with the encryption of TensorFlow. The encrypted algorithm prevents transmission data from being intercepted. Although the transmitted data contains no sensitive information, it is still possible that privacy-related data can be inferred from non-sensitive information.
At the same time, to prevent the identification of individuals from attribute distributions, our model employs multiple strategies like syntactic anonymization models and differential privacy models.
Structure representations are reconstructed by embedding representations, and reconstructed parameters could be used to adjust accuracy.
In fact, a small difference between raw structures and reconstructed structures can protect privacy~\cite{DBLP:journals/tvcg/WangCCBGCPM19}.
In the fields of secure multiple computing and homomorphic encryption, there are various strategies to handle different privacy and security issues. Our approach is fully compatible with them and supports interactive configurations.

\textbf{Expansibility.}
Our approach can accomplish various graph visual analysis tasks. Three components are employed for constructing different types of graph representations: embedding representation, attribute representation, and structure representation. Users can freely configure strategies, methods, and parameters of each component.
As shown in case studies, users accomplished different analysis tasks for multi-party graphs, including anomaly detection, clustering, and comparison.
Experts highly rated our approach in gaining and identifying patterns.
Our approach also supports to design new components for constructing distinctive graph representations. Various visualization styles for different graph representations can also be employed to fulfill a variety of complex tasks.

\textbf{Scalability.}
Our approach constructs federated graph representations from multi-party graphs with reasonable scalability.
\model is compatible with different models and encrypt strategies for different tasks and requirements.
We conducted multiple experiments to measure the performance of the federated graph representation (\autoref{sec-fml-evaluate}).
The efficiency of \model in terms of data size depends on the selected model. For the embedding component, GAT can only handle a moderate-sized data due to the use of matrix, and DeepWalk can support large-sized data because it uses the skip-gram technique.
The other two components also support extracting representations from large-scale graphs.
Our in-house visualization engine is amenable for visualizing large-scaled graphs with rich user interactions.

\textbf{Performance.}
Our model indeed extracted high-quality federated graph representations from multi-party graphs. Federated representations improve the efficiency of anomaly detection and clustering results compared with using the representation extracted from single data.
\name with rich interactions empowers experts to accelerate the process of detecting anomalous trades and comparing clusters of papers.
There are three conclusions drawn from experiments.
\begin{itemize}[parsep=1pt, topsep=3pt]
\item  Compared with the centralized counterpart, our approach can generate results with a similar quality, and achieve a better running performance.
\item The graph can be reconstructed well with federated structure representations in terms of AUC score and precision, and the reconstructed structures keep differences to prevent privacy leaks.
\item The federated attribute representation can be constructed with relatively low computing costs.
\end{itemize}

\section{Conclusion}
This paper presents \name, a federation approach that constructs joint representations of multi-party graphs, and supports privacy-preserving visual analysis of multi-party graphs.
In the future, we plan to explore various encryption strategies. We alsp plan to extend our approach to other graph data. Currently, we assume that multi-party graphs have identical attributes. We will improve \model to support heterogeneous graphs.



\bibliographystyle{abbrv-doi}

\bibliography{ref-0430-zjh.bib}
\end{document}